\documentclass[12pt]{article}
\usepackage{epsfig}
\usepackage{bm}
\newcommand\ba{\begin{eqnarray}}
\newcommand\ea{\end{eqnarray}}

\begin{document}

\begin{titlepage}
\begin{center}

\vspace{5cm}

{\Large{\bf
  Single Spin Asymmetries in
High Energy Reactions and Nonperturbative QCD Effects}}
\vspace{0.50cm}\\
A.E. Dorokhov$^{a}$\footnote{dorokhov@theor.jinr.ru},
 N.I. Kochelev$^{a}$\footnote{kochelev@theor.jinr.ru},
W.-D. Nowak $^b$,\footnote{Wolf-Dieter.Nowakw@ifh.de},
\vskip 1ex {(a) \it Bogoliubov Laboratory of Theoretical Physics,
Joint Institute for Nuclear Research, Dubna, Moscow region,
141980, Russia}\\
\vskip 1ex {(b) \it DESY-Zeuthen, Platanenallee 6, D-15738 Zeuthen, Germany}\\
 \vskip 1ex
\end{center}

\vskip 0.5cm \centerline{\bf Abstract}
We discuss some  experimental and theoretical results on single
spin asymmetries (SSA) in high energy lepton-hadron and hadron-hadron
 reactions.
In particular,  recent results on meson SSA obtained by HERMES
are considered
in detail. We  also discuss the
SSA results obtained recently by COMPASS,
as well as those from BRAHMS, PHENIX and STAR.
Special attention is paid to a possible nonperturbative QCD
 mechanism
that might be responsible for the observed meson SSA.
 This mechanism originates
from the spin-flip quark-gluon chromomagnetic interaction
induced by the complex topological structure of the QCD vacuum.
We argue that 
in semi-inclusive deep-inelastic scattering a large SSA is  expected
not only for mesons but also for  baryons
due to  strong nonperturbative final state interactions  between
$ud$-diquark and $u$-quark in the fragmenting  proton.

\vskip 0.3cm \leftline{Pacs: 24.85.+p, 12.38.-t, 12.38.Mh, 12.39.Mk}
\leftline{Keywords: single spin asymmetry, semi-inclusive production,
 non-perturbative QCD}
\end{titlepage}

\vspace{1cm}

\setcounter{footnote}{0}

\section{ Introduction}

The spin structure of the nucleon remains a very intriguing topic.
 In spite of tremendous experimental and theoretical efforts
in the last twenty years, yet there is no clear understanding of the
 spin structure of the nucleon based on fundamental QCD theory
(see,e.g., the recent review \cite{Burkardt:2008jw}).
One promising  way to progress in understanding  the spin structure of the nucleon is to
 study various single spin asymmetries (SSA) in inclusive meson production in
lepton-hadron and hadron-hadron interactions \cite{murgia,efremov}.
At present,  several experiments are producing very important results on
SSA in high energy reactions. In particular, the HERMES Collaboration at DESY
 has observed significant
asymmetries in pion electroproduction on both longitudinally
\cite{HERMES1}, and transversely polarized proton targets
\cite{HERMES2,HERMESlast}. Furthermore, recently HERMES also announced preliminary
data on the observation of a very  large SSA  for $K^+$ \cite{HERMESlast}.
These data support the idea that SSA do not vanish in the high energy
limit $\sqrt{s}>> \Lambda_{QCD}$, in contrast to the expectation of 
the naive perturbative QCD (pQCD) approach
$A_N\propto\alpha_sm_q/\sqrt{s}\ll 1$, in which SSA should be suppressed 
by both
the small value of the QCD coupling constant $\alpha_s$ and the small  current quark mass $m_q$
\cite{Kane:1978nd}.
We recall that
SSA at high energy and large transverse momenta
were also  found  in inclusive and exclusive
hadron production in hadron-hadron collisions
a long time ago~\cite{E704,Krisch:2007zza}.
 At present, the investigations of SSA in hadron-hadron
collisions are continued by STAR, PHENIX and BRAHMS at RHIC using
very large energies \cite{STAR,BRAHMS}.

In this paper  we will discuss experimental results
 and indicate a
possible explanation of large observed SSA within the nonperturbative QCD
 approach.\\

\section{ SSA results in SIDIS}

The main tool to study the mechanism of SSA in QCD using lepton-nucleon scattering
is the measurement of SSA for $\pi$ and $K$ meson inclusive
electroproduction off a
polarized nucleon target, i.e. semi-inclusive deep-inelastic scattering
(SIDIS). The HERMES Collaboration  \cite{HERMESDESY}
was using the polarized 25.7 GeV $e^+/e^-$ beam of HERA,
 which was scattered on a polarized
proton target. The SSA measurements  at HERMES are  devoted to the extraction of information
on the so-called Sivers distribution and Collins fragmentation functions which are appearing  in the
leading twist approach to SSA based on perturbative QCD (pQCD) factorization
\cite{murgia,efremov}.
These functions carry  important information on
the structure of strong interactions at large distances and can be
calculated only within nonperturbative QCD. Fortunately, in SIDIS on a transversely polarized target
there is the possibility to separate the effects of these two functions
because they induce  two distinct angular dependences.
The SSA for a transversely
polarized target,
\begin{equation}
A_{UT}^h(\phi,\phi_s)=\frac{1}{\mid S _\bot \mid}\frac{N^{\uparrow}_h(\phi,\phi_s)
-N^{\downarrow}_h(\phi,\phi_s)}{N^{\uparrow}_h(\phi,\phi_s)
+N^{\downarrow}_h(\phi,\phi_s)}
\label{ssa}
\end{equation}
 receives contributions from  Sivers
\begin{equation}
A_{UT}^{Siv}(\phi,\phi_s)\propto sin(\phi-\phi_s)
\label{siv}
\end{equation}
and Collins
\begin{equation}
A_{UT}^{Col}(\phi,\phi_s)\propto sin(\phi+\phi_s)
\label{col}
\end{equation}
asymmetries which are proportional to the
corresponding Sivers and Collins functions, respectively.
The definition of angles used  in (\ref{ssa}),(\ref{siv}) and ({\ref{col})
is shown  in Fig.1
\begin{figure}[h]
\centerline{\epsfig{file=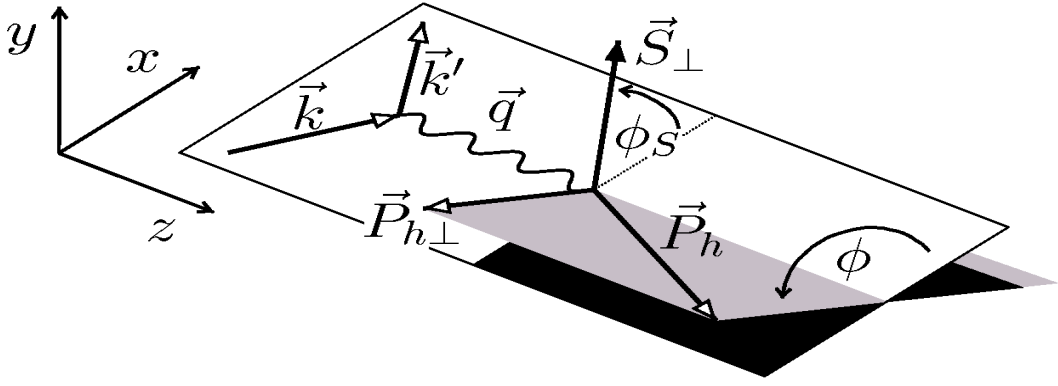,width=14cm, height=5cm,angle=0}}\
\caption{Polarized SIDIS kinematics.}
\label{tree}
\end{figure}
The kinematic range of the HERMES experiment for
the SIDIS reaction
\begin{equation}
e(k)+P(P)\rightarrow e(k^\prime)+h(P_h)+X(P_X)
\end{equation}
is as follows:
\begin{equation}
W^2>10 \  {\rm GeV}^2,\ \  Q^2>1 \ {\rm GeV}^2,\ \  0.1<y<0.85,\ \   0.2<z<0.7,
\end{equation}
where
\begin{equation}
 W^2=(P+q)^2,\ \  Q^2=-q^2=-(k-k^\prime)^2,\ \  y=P\cdot q/(P\cdot k) , \ \ z=P\cdot P_h/(P\cdot q).
\end{equation}
The average value of the Bjorken variable $x=Q^2/(2P.q)$ for the HERMES SSA measurement
 is rather large:
$\langle x\rangle =0.09$. The average value of the proton polarization was $S_T=0.78\pm0.04 $.
In Fig.2 HERMES data on Collins and Sivers asymmetries are presented.
\begin{figure}[h]
\begin{tabular}{cc}
\includegraphics[scale=0.33]{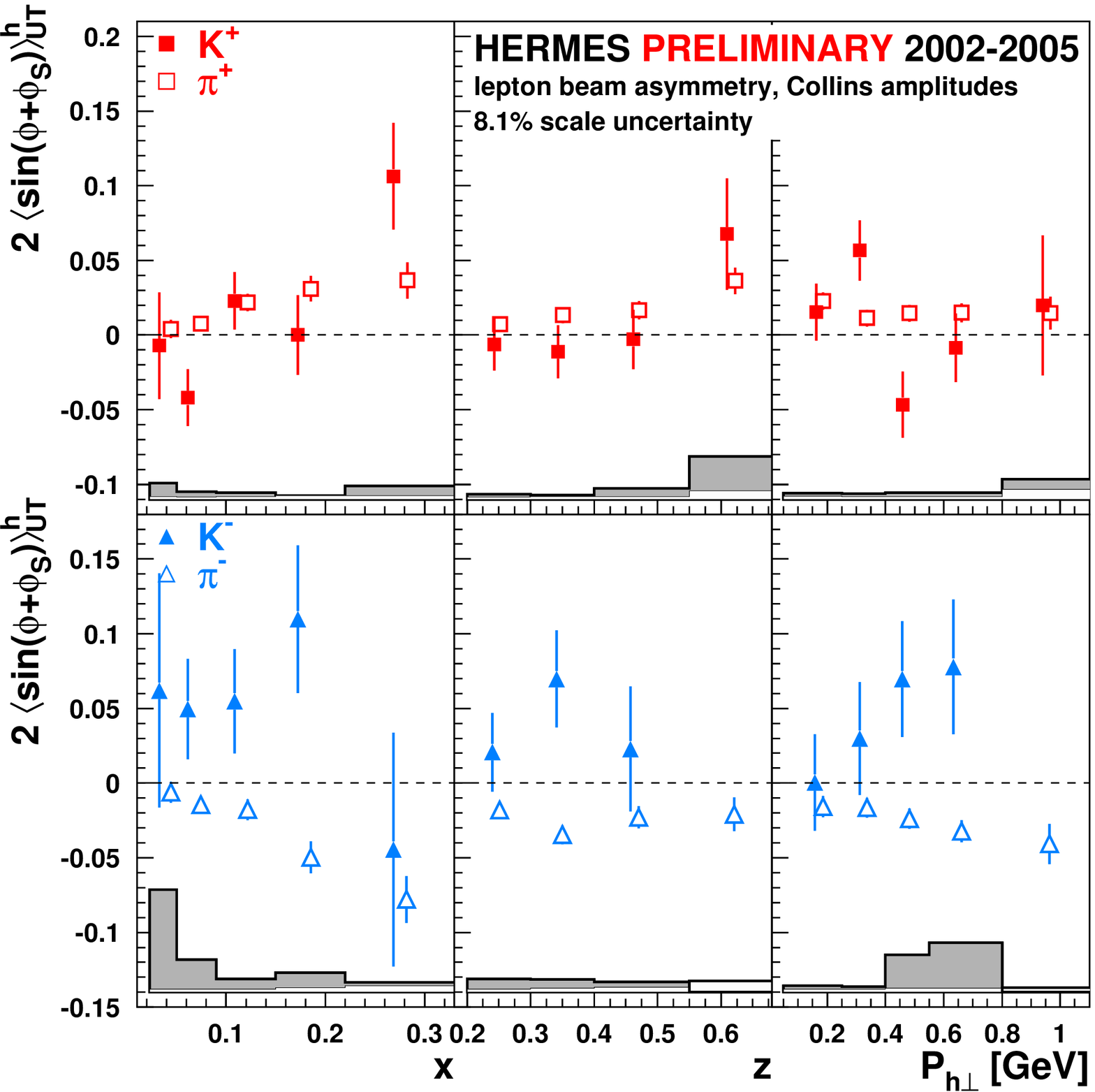} &
\includegraphics[scale=0.33]{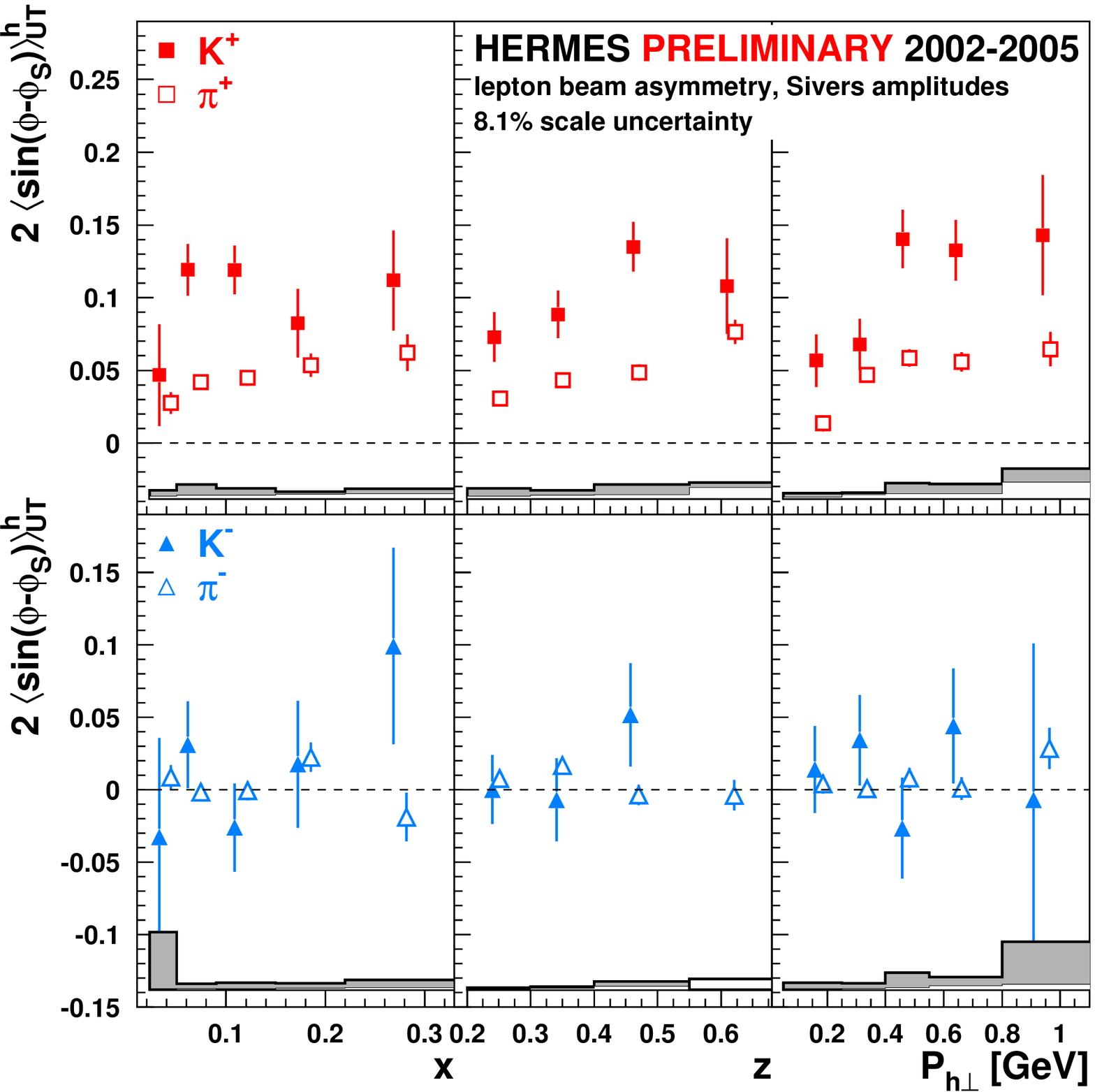}
\end{tabular}
\caption{Collins amplitudes (left panel) and Sivers amplitudes (right
panel) for charged kaons (closed symbols, as labelled) and charged
pions (open symbols, as labelled) as function of $x$, $z$
and $P_{h\bot} $.}
\label{ssa-mesoncomparion}
\end{figure}

The first main feature of the HERMES data is a large Collins and a small Sivers
asymmetries for negatively charged particles.
The second observation is a very large $K^+$ Sivers asymmetry. This asymmetry
is bigger than the $\pi^+$ asymmetry by approximately a factor of three.
Such an anomalous $K^+$ asymmetry is very difficult to explain in the naive pQCD
factorization approach to SSA, in which the main contribution to
both $K^+$ and $\pi^+$ SSA is coming from valence $u$-quark fragmentation.

Recently, first results on SSA in SIDIS
for $\pi$- and $K$-mesons  produced  by scattering positively charged muons
with momentum 160 GeV/c off a deuteron target
\cite{:2008dn},
and for positively and negatively charged particles off
a proton target \cite{Levorato:2008tv}, were presented by the COMPASS
Collaboration \cite{COMPASS}.
The main difference compared to the HERMES  kinematics is the  much smaller value of $x$,
in the range $0.008\div 0.02$.
One of the unexpected results obtained by
 COMPASS  is that practically all SSA in SIDIS
are compatible with zero, with the exception of an indication of a
nonzero Collins asymmetry on the proton for both negative
and positive hadrons at  $x>0.05$.
A possible explanation for the difference in the observed SSA between
HERMES and COMPASS is the rather different kinematic region of SIDIS
explored by them.

\section{SSA in hadron-hadron interactions}

Additional information about the mechanism of SSA is coming from inclusive
particle production in hadron-hadron interactions.
Unfortunately, in this case it is impossible to separate Sivers and Collins
asymmetries because no information is available about the parton scattering
plane.

\begin{figure}[h]
\begin{minipage}[c]{8cm}
\centerline{\epsfig{file=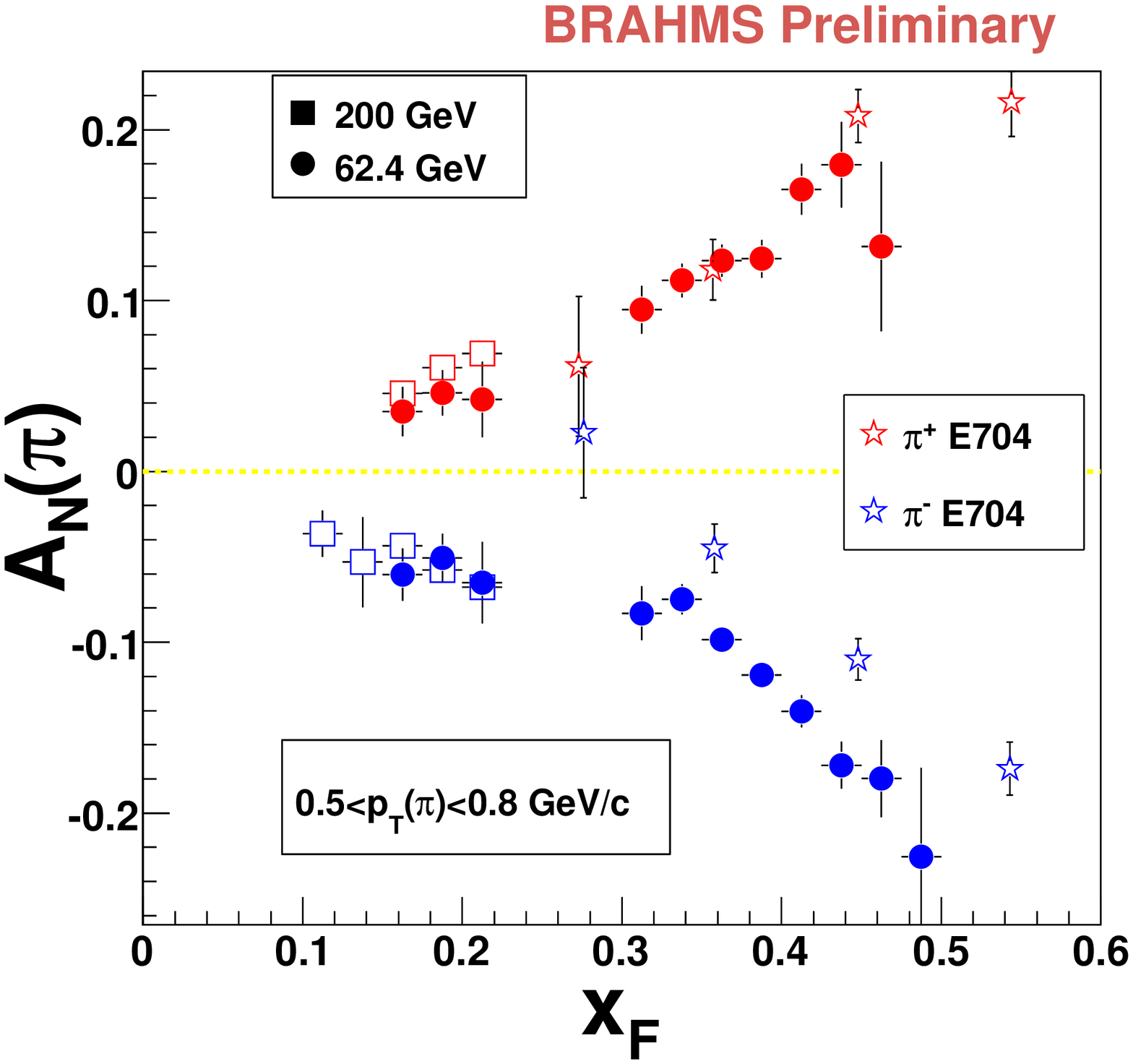,width=7cm, height=5cm,
angle=0}}\
\vspace*{-0.5cm}
\caption{
Comparison of charged pion asymmetries measured at 200 and 62.4~GeV by
BRAHMS \cite{BRAHMS}
and at 19.4~GeV by E704 \cite{E704}.}
\end{minipage}
\hspace*{0.5cm}
\begin{minipage}[c]{8cm}
\vspace*{-0.5cm}
\centering \centerline{\epsfig{file=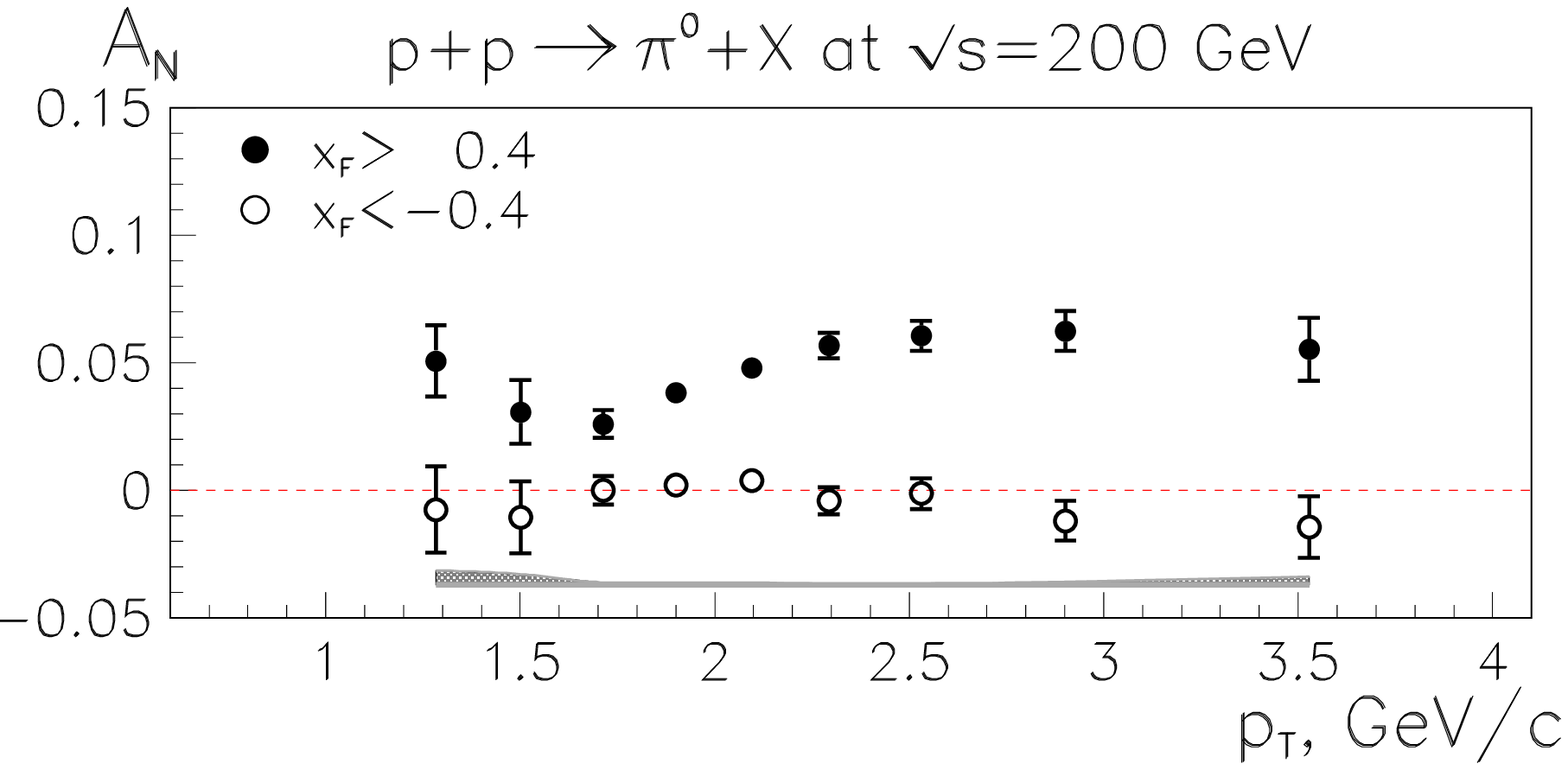,width=6cm, height=5cm,angle=0}}\
\vspace*{-0.5cm}
\caption{STAR experiment SSA data
at $\sqrt{s}=200$ GeV  for neutral pions as a function of $P_{h\bot} $
\cite{STAR}.}
\end{minipage}
\end{figure}
In this situation, one way to describe SSA in hadron-hadron
 scattering is using as input the  Sivers and Collins functions
which were extracted from  fits to SIDIS data. However, this procedure
is based on several strong assumptions. One of them is the assumption
about the validity of factorization in
the description of SSA. Moreover,
the mechanisms of SSA in SIDIS and hadron-hadron scattering might be different
(see discussion below). One evidence here is that SSA in
hadron-hadron scattering practically do not show any energy dependence
(Fig.3), in contrast to SIDIS where there are
rather large differences in the SSA results  obtained by HERMES and COMPASS.
We  would also like to point out that
 hadron-hadron  data on SSA in elastic reactions \cite{Krisch:2007zza}
show some strong oscillations as a function of $p_{h\bot}$. Some evidence
for such oscillations one can also see in Fig.4 where recent STAR data on SSA
for neutral pion production are presented \cite{STAR}.
 Such behavior has never been observed in SIDIS and therefore
 can not be described by using the Sivers and Collins functions extracted
from SIDIS data.

\section{Nonperturbative quark-gluon  interaction as a
source of SSA in high energy reactions}

As it was mentioned in the Introduction, even assuming factorization it is necessary 
to use some nonperturbative input to describe SSA.
There are several calculations of the Sivers function which include
nonperturbative dynamics in different ways.
A first estimate of this function was obtained within the MIT bag model
 \cite{MIT}. Recently, a calculation of the Sivers function was
performed within the Isgur-Karl model \cite{vento}.
 However, we  emphasize
 that these models can be used only for some qualitative estimates because
they are based on the assumption of the dominance of
perturbative gluon exchange between struck and spectator
  quarks.  It is very hard to
expect the validity of such an assumption for  transverse momenta
$p_\bot\leq 1$ GeV.
In this approach, nonperturbative effects in SSA have the scale
$p_\bot \approx 1/R_{conf}\approx \Lambda_{QCD}$ and are related only
to  confinement dynamics but they are not included in the interaction
between struck and spectator quarks. In this respect, the final-state
interaction mechanism for SSA considered in \cite{FSI} also
belongs to the above class of models.

The model of the nonperturbative QCD vacuum
based on strong topological fluctuations of gluon
fields called instantons is one of the most successful models
for nonperturbative QCD effects (see the review \cite{Schafer:1996wv}). It has been shown that
additionally to the famous multiquark t'Hooft interaction, which
gives rise to a  negative sea quark polarization in
the proton \cite{Dorokhov:1993ym},
 instantons also lead to {\it spin-flip quark-gluon
chromomagnetic interactions}, which should have strong effects  to SSA in
high-energy reactions \cite{kochelev2} (see also recent the
discussion in \cite{diakonov} and \cite{Hoyer:2005ev}).
The effect of including such interaction,
\begin{equation}
{\cal L_I}^{chromo}= -i\frac{g_s\mu_a}{2m^*_q} \bar q
\sigma_{\mu\nu}t^aG^a_{\mu\nu}q,
 \label{chrom}
 \end{equation}
  where $ \mu_a$ is the quark anomalous
chromomagnetic moment, on the Sivers function was considered in \cite{ins}.
This contribution is arising from the interference of the diagrams
presented in Fig.5.
\begin{figure}[h]
\centerline{\epsfig{file=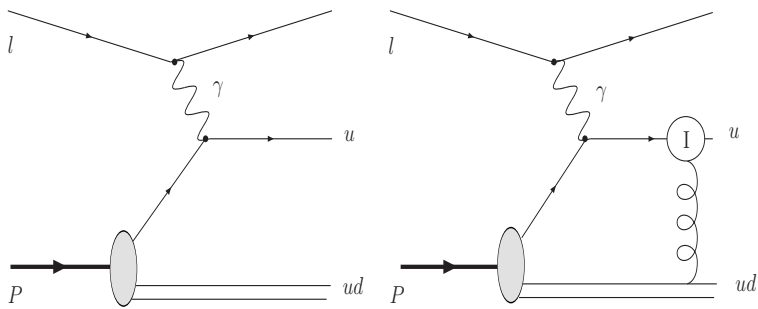,width=4cm, height=10cm,
angle=90}}\
\caption{ The
diagrams  giving rise  to SSA in SIDIS.
The symbol I denotes the instanton.}
\end{figure}

It was shown that this
contribution is very large \cite{ins} and not suppressed by
powers of the strong coupling constant. The specific flavour
dependence of the instanton contribution leads to a large
negative contribution to the $u$-quark Sivers function. Furthermore,
it also leads to a small Sivers function for the $d$-quark. This is in
qualitative agreement with the HERMES result on a large Sivers asymmetry for positively
charged pions and a small one  for negatively charged pions (see Fig.2).
For a more detailed comparison with data one should additionally include
the contribution to SSA
originating from the final-state interaction (FSI) induced by non-perturbative
quark exchange between struck and spectator quarks (Fig.6) \cite{KBK}.

\begin{figure}[h]
\centerline{\epsfig{file=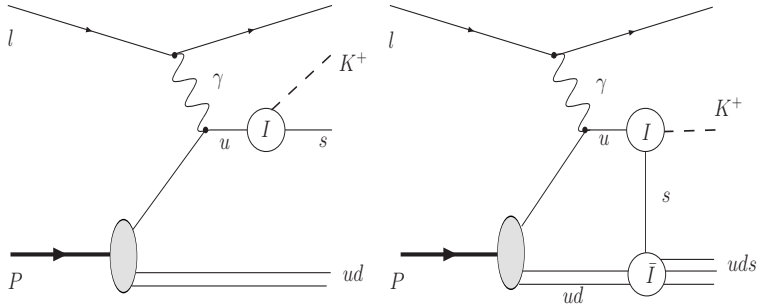,width=4cm,height=10cm, angle=90}}\
 \caption{
The quark-exchange FSI contributing to $K^+$ SSA in SIDIS.}
\end{figure}

\begin{figure}[h]
\centerline{\epsfig{file=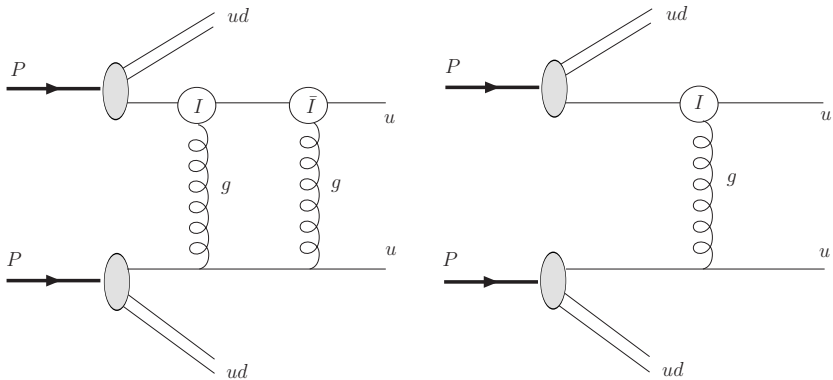,width=5cm,height=11cm, angle=90}}\
 \caption{The diagrams which lead to nonzero SSA in hadron-hadron interaction.}
\end{figure}
Finally, we emphasize  that in fact the
 mechanisms leading to large SSA in SIDIS as presented in  Fig.5
 and for hadron-hadron
interactions \cite{kochelev2} might
be  different, especially in the small $p_\bot$ region.
 As an example, in Fig.7 we present
the diagrams which might be responsible for SSA in hadron-hadron reactions
(such a two-gluon exchange is subleading in SIDIS).

\section{ SSA for baryons in SIDIS}

The  mechanism that
might be responsible for large observed SSA for  different types
of hyperons is   one of the  long-standing problems
in strong interactions \cite{Nurushev:1997sx,Bravar:1995fw}. In particular,
a very large SSA for $\Lambda$-hyperons was observed with both polarized and
unpolarized hadron beams.
Recently, a large SSA for neutrons at large $x_F$ was observed by
the PHENIX Collaboration \cite{Bazilevsky:2006vd} at 
$\sqrt{s}=200$  GeV. Such a large asymmetry is very difficult to explain in
the conventional Regge approach \cite{Kopeliovich:2008hy}.
We  point out that the FSI between $u$-quark and
$ud$-diquark depicted in Fig. 5
should lead not only to a SSA for the  meson coming from $u$-quark
fragmentation
 but also to a SSA for baryons to which a $ud$-diquark should  mainly
 fragment. Indeed, in  general, on the quark level the $u$-quark  SSA in SIDIS is proportional to:
\begin{equation}
A^u=Ae_{\mu\nu\rho\sigma}S_\mu p_\nu q_\rho k_{u},
\label{SSAu}
\end{equation}
 where $A$ is some function of dynamical variables,
$S$ is the proton spin and $p$,$q$ and $k_u$ are the momenta of
proton, photon and $u$-quark, respectively.
Due to  conservation of total momentum one has
\begin{equation}
p+q=k_{u}+k_{ud},
\end{equation}
where $k_{ud}$ is the $ud$-diquark momentum
and, therefore, the  asymmetry for a $ud$-diquark should be opposite in sign to
the $u$-quark asymmetry:
\begin{equation}
A^{ud}=-Ae_{\mu\nu\rho\sigma}S_\mu p_\nu q_\rho k_{ud}.
\label{SSAud}
\end{equation}
From  Eqs. \ref{SSAu} and \ref{SSAud}
it follows that  one should expect a strong correlation between SSA for mesons and
baryons  which are fragmenting from $u$-quark and $ud$-diquark, respectively.
For example, in  the case of  scalar $ud$-diquark
dominance in the proton wave function,  we  predict for the Sivers asymmetry in SIDIS
for directly produced $\Lambda$-hyperons and neutrons \footnote{We do not consider here the possible additional
contributions  to SSA
if these baryons originate from decays of  heavier resonances.} :
\begin{eqnarray}
{A^{siv}}({\Lambda})&\approx& -{A^{siv}}({K^+})   \label{lamda}\\
{A^{siv}}({neutron})&\approx& -{A^{siv}}({\pi^+}).
\label{sivbar}
\end{eqnarray}
In principle, such baryon asymmetries are hence expected to be large at HERMES kinematics
because the
experimental values for $K^+$ and $\pi^+$ Sivers SSA measured by HERMES are
quite large
\cite{HERMES1,HERMES2,HERMESlast}. The predictions in Eqs. \ref{lamda}, \ref{sivbar} 
are mainly based on   momentum conservation and therefore   quite
 general with a weak dependence on  the FSI model used.    
 
Due to large statistics for  $\Lambda$-hyperons collected by HERMES  and COMPASS,
 there may be a real possibility to check our prediction  Eq. \ref{lamda}
in SIDIS for a transversely polarized target.  
For the spin-zero $ud$-diquark the
Collins asymmetry for baryons is expected to be  zero because this asymmetry
is related to the correlation between the spin of the fragmenting quark system and the momentum
of the resulting secondary particles.  
We  also mention that it is difficult to predict
the SSA for the outgoing proton
in the SIDIS reaction $e+p\rightarrow p+X$
because it is well known that at large $x_F$ a large contribution
 to the inclusive cross section is coming from 
diffractive excitation of the proton. This contribution is expected to dilute the SSA
for the final proton.

\section{ Acknowledgments}

We are grateful to  U.~D'~Alesio, I.O. Cherednikov, A.~V.~Efremov,
 S.B. Gerasimov, V.A. Korotkov, E.A. Kuraev, F. Murgia,
 O.V. Teryaev  and   V. Vento
 for  useful discussions.
The  work  was supported by the
Heisenberg-Landau program.

\end{document}